\documentclass[a4paper,11pt]{article}
\pdfoutput=1 % if your are submitting a pdflatex (i.e. if you have
\usepackage{jcappub} % for details on the use of the package, please
\usepackage[T1]{fontenc} % if needed

\title{A Consistent View of Interacting Dark Energy from Multiple CMB Probes}

\author[a]{Yuejia Zhai,}
\author[a,1]{William Giar\`e, \note{Corresponding author.}}
\author[a]{Carsten van de Bruck,}
\author[a]{Eleonora Di Valentino,}
\author[b]{Olga Mena,}
\author[c,d]{Rafael C. Nunes}

\affiliation[a]{School of Mathematics and Statistics, University of Sheffield, Hounsfield Road, Sheffield S3 7RH, United Kingdom}
\affiliation[b]{Instituto de F\'{i}sica Corpuscular (IFIC), University of Valencia-CSIC, Parc Cient\'{i}fic UV, c/ Cate\-dr\'{a}tico Jos\'{e} Beltr\'{a}n 2, E-46980 Paterna, Spain}
\affiliation[c]{Instituto de F\'{i}sica, Universidade Federal do Rio Grande do Sul, 91501-970 Porto Alegre RS, Brazil}
\affiliation[d]{Divis\~{a}o de Astrof\'{i}sica, Instituto Nacional de Pesquisas Espaciais, Avenida dos Astronautas 1758, S\~{a}o Jos\'{e} dos Campos, 12227-010, S\~{a}o Paulo, Brazil}

\emailAdd{yzhai13@sheffield.ac.uk}
\emailAdd{w.giare@sheffield.ac.uk}
\emailAdd{c.vandebruck@sheffield.ac.uk}
\emailAdd{e.divalentino@sheffield.ac.uk}
\emailAdd{omena@ific.uv.es}
\emailAdd{rafadcnunes@gmail.com}

\abstract{We analyze a cosmological model featuring an interaction between dark energy and dark matter in light of the measurements of the Cosmic Microwave Background released by three independent experiments: the most recent data by the Planck satellite and the Atacama Cosmology Telescope, and WMAP (9-year data). We show that different combinations of the datasets provide similar results, always favoring an interacting dark sector with a $95\%$~CL significance in the majority of the cases. Remarkably, such a preference remains consistent when cross-checked through independent probes, while always yielding a value of the expansion rate $H_0$ consistent with the local distance ladder measurements. We investigate the source of this preference by scrutinizing the angular power spectra of temperature and polarization anisotropies as measured by different experiments.}

\begin{document}
\maketitle
\flushbottom

\section{Introduction}
\label{sec:introduction}
The standard cosmological model, known as $\Lambda$CDM, describes the Universe as isotropic and homogeneous on large scales. The majority of the matter in the model is made up of Cold Dark Matter (CDM), which is parametrized as a perfect fluid of collisionless particles that interact solely through gravity. The model also accounts for the existence of Dark Energy, represented by a cosmological constant $\Lambda$ in the Einstein equations, which is responsible for the observed accelerated expansion of the Universe at later stages. To set the initial conditions, the model relies on cosmological inflation, an early phase of almost de-Sitter expansion, which leads the Universe towards homogeneity and flatness, while also providing a compelling explanation for the origin of primordial density fluctuations.

Despite involving such poorly understood physics, the $\Lambda$CDM model has been highly successful over the last few decades in providing an accurate fit to a broad range of cosmological and astrophysical observations. Nevertheless, as error-bars on cosmological parameters began to narrow, different intriguing tensions and anomalies emerged at various statistical levels~\cite{Abdalla:2022yfr,Perivolaropoulos:2021jda,DiValentino:2022fjm}. Currently, the most significant tension is between the Hubble constant ($H_0$) value, as measured by the SH0ES collaboration, that is using a distance ladder with Cepheids variables to calibrate Type Ia supernovae~\cite{Riess:2021jrx} ($H_0=73.04 \pm 1.04$ km/s/Mpc), and the value inferred by the Planck satellite from Cosmic Microwave Background (CMB) observations~\cite{Planck:2018vyg} ($H_0=67.4\pm0.5$ km/s/Mpc) assuming a $\Lambda$CDM model for the expansion history of the Universe. The so-called $H_0$ tension~\cite{Verde:2019ivm,DiValentino:2020zio} has recently overcome the threshold of 5 standard deviations~\cite{Riess:2021jrx,Riess:2022mme}, essentially ruling out the possibility of a statistical fluke. It is also important to note that several alternative observations of the late-time Universe support the SH0ES result, and none of these measurements suggests a value lower than early Universe estimates~\cite{Anderson:2023aga,Napier:2023fof,Shajib:2023uig,Dhawan:2022gac,Tully:2022rbj,deJaeger:2022lit,Pesce:2020xfe,Ward:2022ghz,Wang:2022msf,Garnavich:2022hef,Freedman:2020dne,Huang:2019yhh,Blakeslee:2021rqi}. Additionally, Planck-independent observations of the CMB temperature and polarization anisotropies always predict an expansion rate consistent with Planck and never higher than late-time probes~\cite{Aiola:2020azj,SPT-3G:2022hvq}, assuming a $\Lambda$CDM scenario. Consequently, the $H_0$ tension suggests a discrepancy between our comprehension of the early and late Universe.

Certainly, discrepancies in the observational data may point to the presence of unaccounted-for systematic errors. Although it cannot be ruled out entirely, this possibility is becoming increasingly unlikely, given the extensive analysis performed by the SH0ES collaboration~\cite{Riess:2021jrx} and the distribution of the local measurements, that are made by different teams with different probed and calibration methods: when these measurements are combined together, even removing a few of them, the total tension with the CMB estimate still persists at the $4-6\sigma$ level~\cite{Riess:2019qba,DiValentino:2020vnx}.
More excitingly, the $H_0$ tension can indicate the necessity of new physics, because inferring $H_0$, i.e. the rate at which the Universe is expanding today, from observations of the Cosmic Microwave Background necessarily relies on the cosmological model and its underlying assumptions. In more complex cosmologies beyond $\Lambda$CDM, values of $H_0$ consistent with local distance ladder measurements can be obtained, and numerous potential solutions have been proposed in the literature, see, e.g., Refs.~\cite{DiValentino:2021izs,Knox:2019rjx,Jedamzik:2020zmd,Abdalla:2022yfr,Kamionkowski:2022pkx} for recent reviews. 

One model that has gained popularity for potentially resolving the $H_0$ tension is the interacting dark energy (IDE) scenario, where a non-gravitational interaction between dark matter (DM) and dark energy (DE) is postulated~\cite{Valiviita:2008iv,Gavela:2009cy,DiValentino:2017iww, Kumar:2017dnp, Wang:2016lxa, Martinelli:2019dau, Yang:2019uog, DiValentino:2019ffd, Pan:2019jqh, Kumar:2019wfs, Yang:2018euj, Kumar:2016zpg, Murgia:2016ccp,Pourtsidou:2016ico, Yang:2018ubt, Yang:2019uzo, Pan:2019gop, DiValentino:2019jae, DiValentino:2020leo, Yao:2020pji, Lucca:2020zjb, DiValentino:2020kpf, Gomez-Valent:2020mqn, Yang:2020uga, Yao:2020hkw, Pan:2020bur, DiValentino:2020vnx, Lucca:2021dxo, Kumar:2021eev, Yang:2021hxg, Gao:2021xnk, Yang:2021oxc, Lucca:2021eqy, Halder:2021jiv, Kaneta:2022kjj, Gariazzo:2021qtg, Nunes:2021zzi, Yang:2022csz, Nunes:2022bhn,vanderWesthuizen:2023hcl}. 
The state-of-the-art constraints on IDE cosmologies arise
primarily from the CMB data released by the Planck collaboration, which provides a mild-to-moderate indication for an interacting dark sector, yielding a value of the expansion rate $H_0$ consistent with the local SH0ES measurements. However, when the Planck observations are considered in combination with robust external probes, as measurements of the late-time expansion history from BAO and SN, such preference is usually mitigated~\cite{Nunes:2022bhn}. Nonetheless, recent studies suggest a significant reliance on the cosmological model in the matter clustering 3D BAO measurements~\cite{Bernui:2023byc}. Exploiting a 2D transverse projection of the BAO dataset (where the dependence on the cosmological model is much reduced) leads to very different constraints than the traditional 3D BAO approach, resulting in a  very strong evidence in favor of IDE cosmologies.

In this paper, we aim to evaluate the robustness of this indication for Interacting Dark Energy by further testing it against different CMB observations, beyond those from the  Planck satellite~\cite{Planck:2018vyg}. Namely, we explore the constraints derived from the Atacama Cosmology Telescope data~\citep{ACT:2020frw}, both alone and in combination with the 9-year data release from the WMAP satellite~\cite{Hinshaw:2012aka}. Our findings reveal that different independent combinations of data yield comparable results in favor of the IDE scenario, which is consistent with the bounds  obtained from the Planck experiment solely. To understand the underlying cause of this preference, we conduct a comprehensive analysis of the angular power spectra of temperature anisotropies and polarization as measured by the different CMB probes across various scales.

This work is organized as follows: in \autoref{sec:theory} we describe the theoretical model. In \autoref{sec:data} we discuss the datasets and the methodology used in our analyses. In \autoref{sec:res} we discuss our results. Finally, in \autoref{sec:conclusions} we present our conclusions.

\section{The IDE model}
\label{sec:theory}

Within the minimal $\Lambda$CDM framework, dark energy and dark matter only interact through gravity. Therefore, due to the energy-momentum conservation, $\nabla_\mu {T_c^\mu}_\nu=\nabla_\mu {T_x^\mu}_\nu=0$, where $c$ and $x$ denote dark matter and dark energy, respectively.
However, there is no a priori reason why these two quantities should not interact in other ways, and it has been shown that an interaction that assumes an energy flow from dark energy to dark matter (or vice versa) is consistent with the data~\cite{Bertolami:2007tq,Bertolami:2007zm,Abdalla:2009mt,Bernui:2023byc}. In the following, we concentrate on an IDE model in which the interaction is featured by the energy density of dark energy $\rho_x$ and 4-velocity of dark matter $v_c$. In the synchronous gauge, the metric is defined as: 
\begin{equation}
    ds^2=a^2 \Big[ -d\tau^2 + \big(\delta_{ij} + h_{ij}\big)dx^idx^j\Big].
\end{equation}

This IDE model introduces energy-momentum transfer from dark energy to dark matter by modifying their individual energy conservation equations as follows:
\begin{align}
    \nabla_\mu {T_c^\mu}_\nu&=+\frac{Q(v_c)_{\nu}}{a}\\
    \nabla_\mu {T_x^\mu}_\nu&=-\frac{Q(v_c)_{\nu}}{a}~.
\end{align}
The energy density transfer rate $Q$ can have many different phenomenological expressions. In this work we focus on an interacting model with an interacting rate given by:
\begin{equation}
    Q=\xi \mathcal{H}\rho_x~,
    \label{eq:xi}
\end{equation}
where $\xi$ is a dimensionless coupling constant, with $(v_c)_{\nu}=a(-1,(v_c)_i)$ in the synchronous gauge. Note that if $\xi<0$ the energy flows from the dark matter sector to the dark energy one.

The energy density perturbation is $\delta=\delta\rho/\rho$ and the divergence of the fluid proper velocity is $\theta=ik\cdot v$. At linear order,  the perturbations in the dark fluids $\delta_{x,c}$ and $\theta_{x,c}$ evolve as~\cite{Valiviita:2008iv,Gavela:2010tm}: 
\begin{align}
    \delta'_x=&-(1+w)\big(\theta_x+\frac{h'}{2}\big)-\xi\big(\frac{kv_T}{3}+\frac{h'}{6}\big) \nonumber \\
             &-3\mathcal{H}(1-w)\Big[\delta_x+\frac{\mathcal{H}\theta_x}{k^2}\big(3(1+w)+\xi\big)\Big],\\
    \delta'_c=&-\theta_c-\frac{1}{2}h'+\xi\mathcal{H}\frac{\rho_x}{\rho_c}(\delta_x-\delta_c)+\xi\frac{\rho_x}{\rho_c}\big(\frac{kv_T}{3}+\frac{h'}{6}\big),\\
    \theta'_x=&2\mathcal{H}\theta_x + \frac{k^2\delta_x}{w+1}+2\mathcal{H}\frac{\xi}{w+1}\theta_x-\xi\mathcal{H}\frac{\theta_c}{w+1},\\
    \theta'_c=&-\mathcal{H}\theta_c,
\end{align}
where $h$ is the trace of metric perturbation $h_{ij}$, and $'$ denotes taking derivative with respect to $\tau$: $h'\equiv\partial h/\partial \tau$. $v_T$ is the centre of mass velocity for the total fluid, defined as:
\begin{equation}
    v_T(k)=\frac{\sum_i (1+w_i)\rho_i \theta_j /k}{\sum_i (\rho_i + P_i)}~,
\end{equation}
where the index $i$ runs from corresponding species of the fluid, here dark matter and dark energy. The sound speed in dark energy rest frame is assumed to be $c_s^2=1$. In this work, $\delta P_x/\delta\rho_x$ in the synchronous gauge is calculated following the discussion in Refs~\cite{Gavela:2009cy,LopezHonorez:2010esq}.

In order to avoid gravitational and 
early-time instabilities we have to impose $w_x \neq -1$ (fixing $w_x=-0.999$) and that $\xi$ and $(1+w_x)$ have opposite signs~\cite{Gavela:2009cy,Gavela:2010tm}. We therefore analyze the $\xi<0$ case, that has also been shown to be able to help with the $H_0$ tension.

\section{Datasets and Methodology}
\label{sec:data}

We exploit the publicly available code \texttt{COBAYA}\cite{Torrado:2020xyz} to study the observational constraints on the IDE cosmological model. The code explores the posterior distributions of a given parameter space using the Markov Chain Monte Carlo (MCMC) sampler developed for \texttt{CosmoMC}~\cite{Lewis:2002ah} and tailored for parameter spaces with speed hierarchy implementing the “fast dragging” procedure~\cite{Neal:2005}. To compute the theoretical model and introduce the possibility of interactions between dark energy and dark matter, we exploit a modified version of the Cosmic Linear Anisotropy Solving System code, \texttt{CLASS}~\cite{Blas:2011rf}. Our baseline sampling parameters are the usual six $\Lambda$CDM parameters, namely the baryon $\omega_{\rm b}\doteq \Omega_{\rm b}h^2$ and cold dark matter $\omega_{\rm c}\doteq\Omega_{\rm c}h^2$ energy densities, the angular size of the horizon at the last scattering surface $\theta_{\rm{MC}}$, the optical depth $\tau$, the amplitude of primordial scalar perturbation $\log(10^{10}A_{\rm s})$ and the scalar spectral index $n_s$.  In addition, we consider the coupling parameter $\xi$ defined in Eq.~(\ref{eq:xi}). We select uniform prior distributions for all the parameters considered in our analysis, except for the optical depth at reionization ($\tau$), for which we adopt a prior distribution that aligns with the CMB dataset, as discussed below. To test the convergence of the chains obtained using this approach, we utilize the Gelman-Rubin criterion~\cite{Gelman:1992zz}, and we establish a threshold for chain convergence of $R-1 \lesssim 0.02$. 

Our baseline CMB datasets consist of:

\begin{itemize}

\item The full \emph{Planck 2018} temperature and polarization likelihood~\cite{Planck:2019nip,Planck:2018vyg,Planck:2018nkj}, in combination with the Planck 2018 lensing likelihood~\citep{Planck:2018lbu}, reconstructed from measurements of the power spectrum of the lensing potential. We refer to this dataset as “Planck”. 

\item The full \emph{Atacama Cosmology Telescope} temperature and polarization DR4 likelihood~\citep{ACT:2020frw}, assuming a conservative Gaussian prior on $\tau=0.065\pm0.015$ as done in~\cite{Aiola:2020azj}. We refer to this dataset as "ACT”. 

\item The full \emph{Atacama Cosmology Telescope} DR4 likelihood, combined with \emph{WMAP} 9-years observations data~\cite{Hinshaw:2012aka} and a Gaussian prior on $\tau = 0.065 \pm 0.015$, as done in~\cite{Aiola:2020azj}. We refer to this dataset combination as "ACT+WMAP."

\item  The full \emph{Atacama Cosmology Telescope} temperature and polarization DR4 likelihood~\citep{ACT:2020frw}, in combination with the \emph{Planck 2018 TT TE EE} likelihood~\cite{Planck:2019nip,Planck:2018vyg,Planck:2018nkj} in the multipole range $2\le \ell \le 650$ and the Planck 2018 lensing likelihood~\citep{Planck:2018lbu}. We refer to this dataset as "ACT+Planck”.

\item A gaussian prior $H_0 = (73.04 \pm 1.04)$ km/s/Mpc on the Hubble constant as measured by the SH0ES collaboration~\cite{Riess:2021jrx}.  We refer to this data set as SH0ES.

\end{itemize}

Finally, to conduct a model comparison, we calculate the Bayesian evidence for each one and then estimate the corresponding Bayes factors, which are normalized to a baseline $\Lambda$CDM scenario (\textit{i.e.}, without an interacting Dark sector). To perform this task, we employ the \texttt{MCEvidence} package, which is publicly available~\cite{Heavens:2017hkr,Heavens:2017afc}\footnote{The \texttt{MCEvidence} package can be accessed at the following link: \url{https://github.com/yabebalFantaye/MCEvidence}.}. This package has been appropriately modified to be compatible with \texttt{COBAYA}. We use the convention of a negative value if the IDE model is preferred against the $\Lambda$CDM scenario, or vice versa, and we refer to the revised Jeffrey's scale by Trotta~\cite{Kass:1995loi,Trotta:2008qt}, to interpret the results. We will say that the evidence is inconclusive if $0 \leq | \ln B_{ij}|  < 1$, weak if $1 \leq | \ln B_{ij}|  < 2.5$, moderate if $2.5 \leq | \ln B_{ij}|  < 5$, strong if $5 \leq | \ln B_{ij}|  < 10$, and very strong if $| \ln B_{ij} | \geq 10$.

%==============================================

\section{Results}
\label{sec:res}

\begin{table*}
\begin{center}
\renewcommand{\arraystretch}{1.5}
\resizebox{0.8 \textwidth}{!}{
\begin{tabular}{l c c c c c c c c c c c c c c c }
\hline
\textbf{Parameter} & \textbf{ Planck } & \textbf{ ACT } & \textbf{ ACT+WMAP } &  \textbf{ ACT+Planck } \\ 
\hline\hline

$ \Omega_\mathrm{b} h^2  $ & $  0.02237\pm 0.00015 $ & $  0.02153\pm 0.00032 $ & $  0.02238\pm 0.00020 $ &  $  0.02238\pm 0.00013 $ \\ 
$ \Omega_\mathrm{c} h^2  $ & $  0.067^{+0.042}_{-0.031}\, (< 0.115 ) $ & $ <0.0754\, (< 0.111) $ & $  0.070^{+0.046}_{-0.021}\, (< 0.117) $ & $  0.067^{+0.042}_{-0.030}\, (< 0.115 ) $ \\ 
$ H_0  $ & $  71.6\pm 2.1 $ & $  72.6^{+3.4}_{-2.6} $ & $  71.3^{+2.6}_{-3.2} $ & $  71.4^{+2.5}_{-2.8} $ \\ 
$ \tau_\mathrm{reio}  $ & $  0.0534\pm 0.0079 $ &$  0.063\pm 0.015 $ & $  0.061\pm 0.014 $ &  $  0.0533\pm 0.0073 $ \\ 
$ \log(10^{10} A_\mathrm{s})  $ & $  3.042\pm 0.016 $ &$  3.046\pm 0.030 $ & $  3.064\pm 0.028 $ &  $  3.047\pm 0.014 $ \\ 
$ n_\mathrm{s}  $ & $  0.9655\pm 0.0045 $ & $  1.010\pm 0.016 $ & $  0.9741_{-0.0064}^{+ 0.0066} $ & $  0.9699\pm 0.0038 $ \\ 
$ \xi  $ & $  -0.40^{+0.23}_{-0.20} $ & $  -0.46^{+0.20}_{-0.28} $ & $ -0.38_{-0.14}^{+0.35} $ & $  -0.40^{+0.27}_{-0.23} $ \\ 
$S_8$ &$1.10^{+0.19}_{-0.35}$ &$1.18^{+0.26}_{-0.38}$ &$1.08^{+0.19}_{-0.31}$ &$1.09^{+0.19}_{-0.34}$\\
\hline
$\ln \mathcal{B}_{ij}$ & $-0.17$ & $-0.07$ & $0.06$ & $-0.25$\\
\hline \hline
\end{tabular} }
\end{center}
\caption{
\small Constraints (upper limits) at 68\% (95\%) CL on the parameters of the IDE model obtained from different combinations of CMB data, without introducing any prior from the SH0ES collaboration. The Bayes factors $\ln \mathcal{B}_{ij} = \ln \mathcal{Z}_{\rm LCDM} - \ln \mathcal{Z}_{\rm IDE}$ calculated as the difference between the evidence for $\Lambda$CDM and IDE model in such a way that a negative value indicates a preference for the IDE model over the $\Lambda$CDM scenario.
}
\label{tab.results}
\end{table*}

We show in \autoref{tab.results} the constraints at 68\%~CL (upper limits at 95\%~CL) on the cosmological parameters for the IDE scenario studied in this work. 

The first and most important thing to notice is that, independently on the CMB data analysed, a coupling between DM and DE $\xi$ is always preferred with a statistical significance above 1$\sigma$. Indeed, in the majority of the cases the preference is very close to the 95\%~CL. The non-zero preferred value of the coupling is translated into a smaller amount of cold dark matter at present, regardless of the CMB observations exploited in the data analysis. Such a lack of cold dark matter is a straightforward consequence of the non-gravitational interaction among the dark sectors: in the presence of a negative coupling in the rate given by Eq.~(\ref{eq:xi}), due to the energy flow among the dark sectors, the current amount of cold dark matter is reduced with respect to the canonical $\Lambda$CDM scenario. The smaller amount of cold dark matter in interacting scenarios is translated into a value of the Hubble constant $H_0$ higher than in standard scenarios, required to compensate for the lower value $\rho_c$. This is a very important outcome of our numerical analyses: \emph{for all the CMB data sets considered here, the mean value of $H_0$ is much larger, and the significance of the $H_0$ tension is therefore strongly reduced}. Notice in addition that the solution to the tension is mainly led by the shift in the mean value of the Hubble constant and not by the larger size of the errors. Interestingly, the model-comparison results lead to a negative values of the Bayes factor $\ln \mathcal{B}_{ij}$ for \emph{most} of the CMB data combinations considered here. Therefore, even if the preference remains inconclusive, there is a tendency from current CMB measurements towards an IDE cosmology. Such a preference could potentially improve with future CMB observations.~\footnote{The rest of the cosmological parameters depicted in \autoref{tab.results}  show values close to those obtained within the $\Lambda$CDM minimal cosmology except for the scalar spectral index, which is close to unity when ACT observations are included in the analyses. We refer the interested reader to the recent work of Ref.~\cite{Giare:2022rvg} for a comprehensive analysis.}

\begin{figure*}
         \centering
        \includegraphics[width=\textwidth]{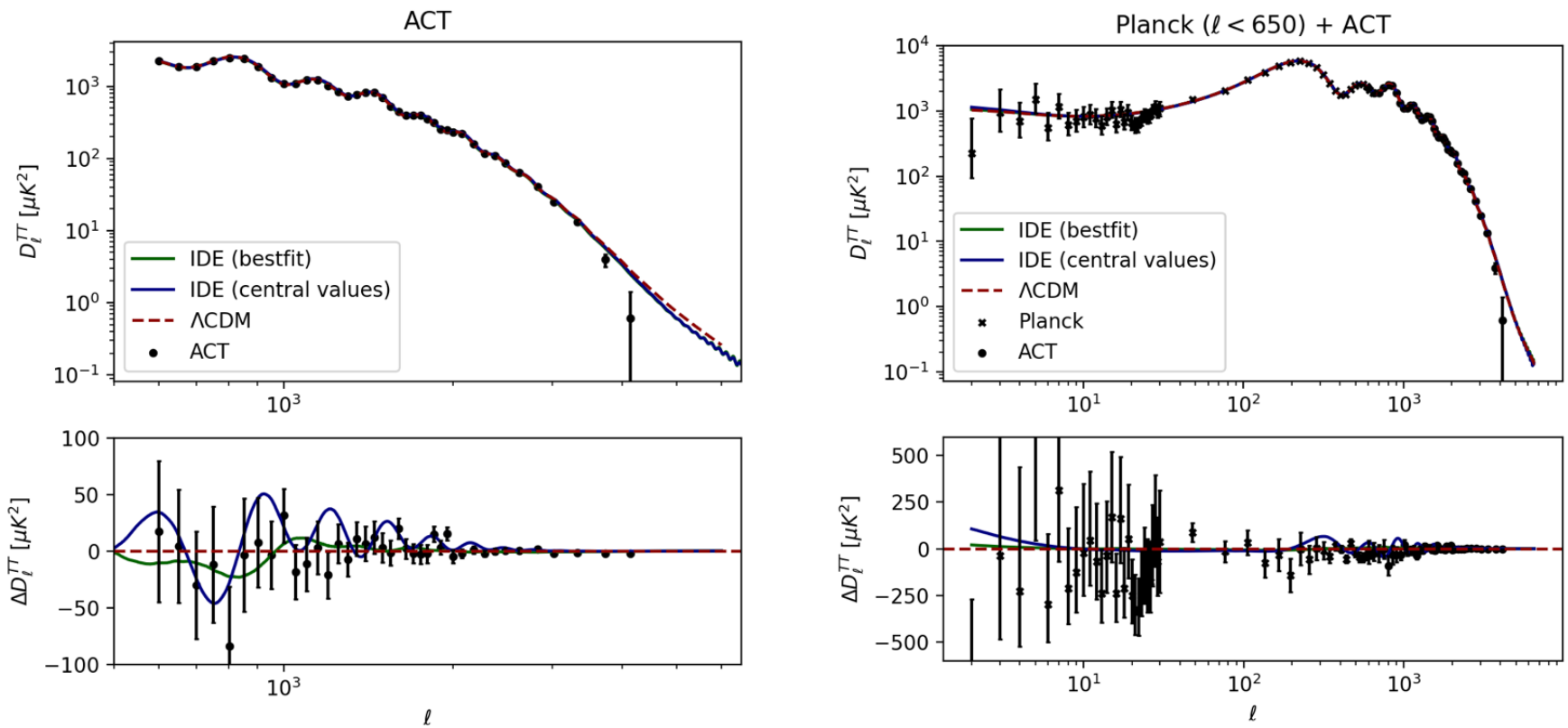}
        
	\caption{ \small Left (right) panel: Comparison of the ACT (Planck plus ACT) best fit and mean-value temperature angular power spectra (upper panels) for IDE and $\Lambda$CDM  cosmologies and residuals (lower panels), plotted against the ACT and Planck ($\ell<650$) data.}
 
        \label{fig:residuals}
\end{figure*}

\autoref{fig:residuals} depicts a comparison of the ACT and Planck plus ACT  results for the temperature angular power spectra in the upper panels, together with the residuals (showing the departure from the minimal $\Lambda$CDM cosmology) in the lower panels. In the case of Planck data we only show the low multipole measurements ($\ell<650$) data. Data points for these two experiments are also illustrated. We show the results for both the canonical $\Lambda$CDM as well as for IDE cosmologies from the best-fit and mean-value cosmological parameters arising from our Monte Carlo data analyses. Focusing on the multipole range $2\lesssim\ell\lesssim 650$ probed by the Planck data, the most notable difference in the angular power spectra is observed at very low multipoles\footnote{Notice that the IDE effects enhance power in the lowest multipoles, known as the ISW plateau. The amplitude of the ISW plateau is primarily controlled by the spectral index $n_s$ and the amplitude of the late-time ISW effect which depends on the duration of the dark energy-dominated era, approximately given by the ratio $\Omega_{\rm{DE}}/\Omega_{m} \simeq \Omega_{\rm{DE}}/(1-\Omega_{\rm{DE}})$. While $n_s$ is almost indistinguishable between the IDE and $\Lambda$CDM, the IDE model predicts a transfer of energy from dark matter to dark energy, resulting in a higher value of $\Omega_{\rm{DE}}$. A larger $\Omega_{\rm{DE}}$ implies a longer period of DE domination and, consequently, an enhanced late-time ISW effect.}  where the error bars are significantly large and at $\ell \sim 300$ where the model exhibits a slight deviation from the observed data-points (also leading to a minor worsening of the fit in this specific range of multipoles). Therefore, when combining the two experiments, the high multipole ACT CMB data at $\ell\gtrsim 650$ are mainly those driving the preference for $\xi<0$ and are also responsible for the global improvement in the fit within the context of IDE models with respect to the minimal $\Lambda$CDM. Such an improvement is due to the contribution from the multipole range $650\lesssim\ell\lesssim 1000$, as well as to the lower amplitude of the ACT acoustic modes at high $\ell\gtrsim 3000$. Therefore it is a \emph{real} effect, rather than that explored in Ref.~\cite{DiValentino:2020leo} for Planck data only, in which the detection of a coupling $\xi<0$ was indeed a \emph{fake} effect induced by parameter degeneracies. To further reassess these findings we have combined ACT with the SH0ES prior on the Hubble constant, obtaining a value for the coupling and $H_0$ of $\xi=-0.45_{-0.20}^{+0.24}$ and $H_0=72.9\pm1.1$ respectively. Notice that these values are very close to those obtained with ACT only data (see \autoref{tab.results}) and that the most relevant effect when adding the prior on the Hubble constant is a decrease on its error, being the change on its mean value completely negligible and therefore making the hints for an IDE cosmology from high multipole data a neat result.

\begin{figure*}
	\centering
	\includegraphics[width=\textwidth]{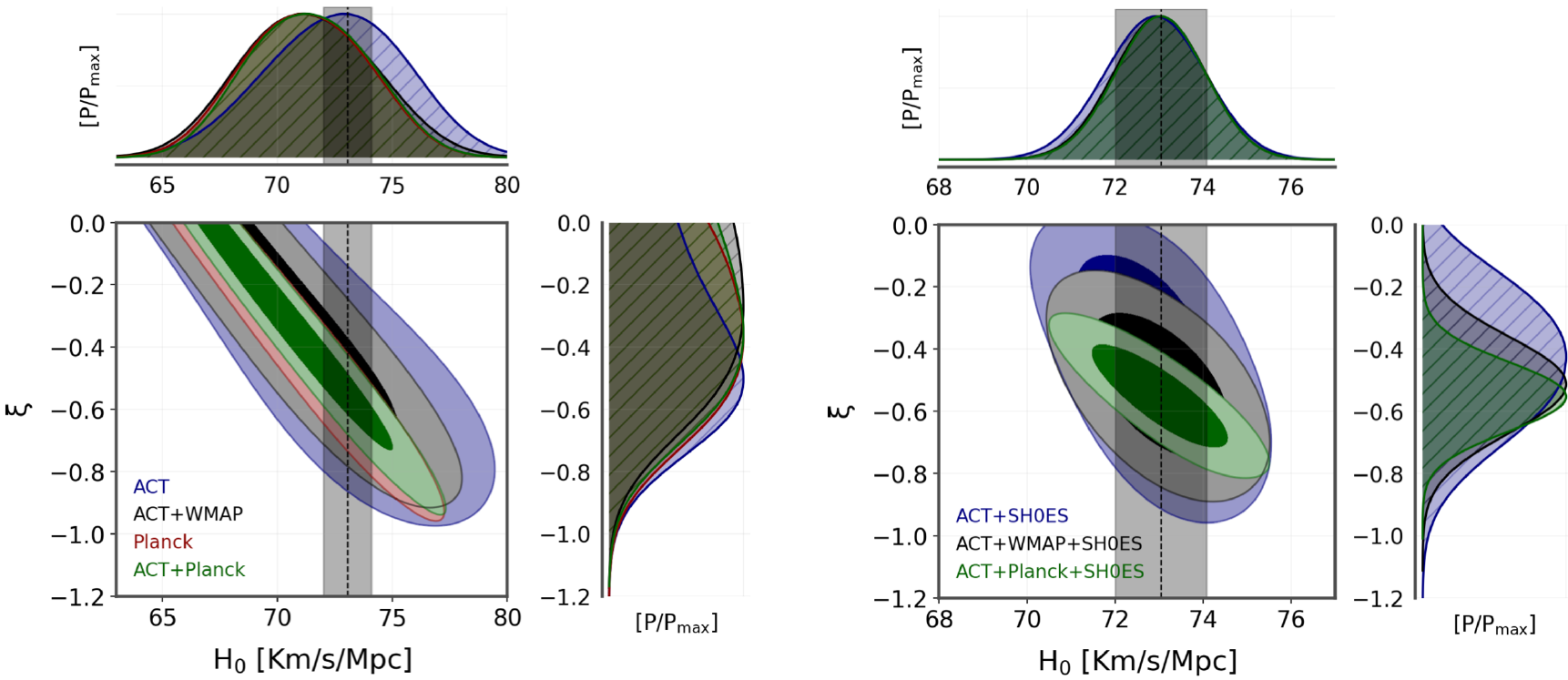}
	\caption{\small 2D contours at 68\% and 95\% CL and 1D posteriors for the coupling parameter $\xi$ and the expansion rate $H_0$, as inferred by the different combinations of CMB data listed in the legend; both  with (right panel) and without (left panel) assuming a prior on the value of $H_0=73.04 \pm 1.04$ km/s/Mpc as measured by the SH0ES collaboration (grey vertical region in the plot).}
\label{fig:coupling}
\end{figure*}

We measure the level of agreement between Planck and ACT under the Interacting Dark Energy cosmology, adopting the Suspiciousness statistics~\cite{Handley:2019wlz,Handley:2020hdp,DiValentino:2022rdg}. This methodology provides a comprehensive overview of how these discrepancies evolve in extended parameter-spaces, without being influenced by biases resulting from prior volume effects. In particular, to avoid any unintended influence of the prior volume, we separate the Bayes Ratio into two parts: the Information ($I$), which is dependent on the prior, and the Suspiciousness ($S$), which is independent of the prior. If the datasets are uncorrelated and the posterior distributions follow a Gaussian-like distribution with means of $\mu$ and a covariance matrix of $\Sigma$, the Suspiciousness can be estimated as~\cite{Handley:2019wlz,Handley:2020hdp,DiValentino:2022rdg}
\begin{equation}
\log S=\frac{d}{2}-\frac{\chi^{2}}{2}
\label{eq:logS}
\end{equation}
where $d$ represents the dimension of the parameter volume of the cosmological model and $\chi^2$ is given by
\begin{equation}
\chi^{2}=\left(\mu_{A}-\mu_{B}\right)\left(\Sigma_{A}+\Sigma_{B}\right)^{-1}\left(\mu_{A}-\mu_{B}\right)
\label{eq:chi2}
\end{equation}
with $[A\,,\,B]\equiv [\rm{Planck}\,,\,\rm{ACT}]$. Notice that the $\chi^2$ can be converted easily into a tension probability by the survival function of the $\chi^2$ distribution
\begin{equation}
p=\int_{\chi^{2}}^{\infty} \frac{x^{d / 2-1} e^{-x / 2}}{2^{d / 2} \Gamma(d / 2)} d x
\label{eq:p}
\end{equation}
and, so into a Gaussian equivalent tension via the inverse error function:
\begin{equation}
\sigma(p)=\sqrt{2} \operatorname{erfc}^{-1}(1-p).
\label{eq:sigma}
\end{equation} 
For the 7-dimensional IDE model considered in this work, we find that the tension between ACT and Planck persists at a Gaussian equivalent level of $2.3\sigma$ ($\log S= -4.7$, $p=0.0217$, $\chi^2=16.4$ ), similar to the baseline cosmological model~\cite{Handley:2020hdp,DiValentino:2022rdg}. Therefore, we conclude that IDE cosmologies do not provide a resolution to the discrepancies observed between the two CMB probes. In fact, the tension between ACT and Planck is mainly driven by differences in the value of the spectral index $n_s$ and $\Omega_b h^2$~\cite{Aiola:2020azj,Giare:2022rvg}, and both present also in this model, as evident from \autoref{tab.results}.

We would like to conclude this section with some final remarks: the substantial flow of energy (of approximately 40\%) from the dark matter sector to the dark energy sector consistently favored by all the CMB probes analyzed in this study, may have significant implications for other cosmological and astrophysical observables. For instance, it can affect the growth of structures in the Universe as well as the fraction of gas $f_{\rm gas}$ derived from galaxy cluster data. In the context of the IDE model, it has been pointed out several times that matter cluster parameters such as $\sigma_8$ or $S_8$ tend to be higher compared to the estimates obtained from weak lensing and galaxy clustering surveys assuming the $\Lambda$CDM framework~\cite{KiDS:2020suj}, see \textit{e.g.}, \cite{DiValentino:2019ffd,Gariazzo:2021qtg,Bernui:2023byc} and discussions therein. However, these estimates also come with large error bars, making it inconclusive to draw any definitive conclusions about the galaxy clustering predictions for IDE models, see also \autoref{tab.results}. Furthermore, it is important to note that the comparison of the these cluster parameters may not be directly relevant as the value of $S_8$ is model-dependent. To properly assess the compatibility, the $S_8$ value derived from weak lensing and galaxy clustering surveys should be compared with values obtained from CMB data, such as Planck or ACT, assuming the same underlying model of cosmology. Similar conclusions can be derived for the determination of $f_{\rm gas}$. Using the most recent measurements available in the literature~\cite{SPT:2021vsu}, it is evident that the $f_{\rm gas}$ fraction is inversely proportional to the total matter density, $\Omega_{m}$, and directly proportional to the Hubble function $H(z)$. In the IDE model, it is predicted that $\Omega_{m}$ is lower at late times compared to $\Lambda$CDM while the expansion rate of the Universe $H(z)$ is expected to be higher. As a result, one can argue that the IDE model would predict a higher $f_{\rm gas}$ compared to $\Lambda$CDM. However, it is essential to highlight once again that accurate modeling of the IDE framework is missing, particularly in terms of its dynamics on non-linear scales and its impact on related observables. Therefore, it is currently premature to draw conclusive findings regarding the tension in the $S_8$ parameter with weak lensing and galaxy clustering data or other astrophysical observables within the IDE framework. In this regard, future investigations are needed and this is beyond the aim of this work.

\section{Conclusions}
\label{sec:conclusions}

Interacting dark sector cosmologies with an energy-momentum transfer between dark energy and dark matter are very appealing scenarios to be confronted against observations, given the fact that no fundamental symmetry in nature forbids those non-gravitational couplings. In this manuscript we test these cosmologies in light of Cosmic Microwave Background measurements released from three independent experiments: the Planck satellite, the Atacama Cosmology Telescope, and WMAP (9-year data). 

Our results, summarized in \autoref{tab.results} and \autoref{fig:coupling}, point to a preference for a non-zero interacting rate among dark energy and dark matter with a $95\%$~CL significance in the majority of cases. This preference does not depend on the data set nor in the data combination considered, and as a byproduct implies a much higher value for the Hubble constant, which becomes always in agreement with local distance ladder measurements. The reason for this higher value of $H_0$ is due to the transfer of energy from the dark matter sector to the dark energy one, which results in a reduced amount of cold dark matter at the present time, regardless of the CMB observations used in the analyses. This reduction in cold dark matter results from the non-gravitational interaction between the dark sectors, which is featured by a negative coupling in the rate given by Eq.~(\ref{eq:xi}).  To compensate the lower value of $\rho_c$, the mean value of $H_0$ gets significantly higher, and the tension associated with $H_0$ is greatly reduced. Therefore, the model can provide a potential solution to the $H_0$ tension that is primarily driven by this \emph{physical} shift and not by larger errors. By including a prior on $H_0$ based on the value measured by the SH0ES collaboration, the mean value of the Hubble constant barely changes, clearly stating our arguments above. This is evident in the right panel of \autoref{fig:coupling}.

We examined the reason for this preference by analyzing angular power spectra from various experiments. In \autoref{fig:residuals}, the top panels compare the temperature angular power spectra from two experiments, ACT and Planck plus ACT, along with the respective data points. The lower panels illustrate the deviations from the minimal $\Lambda$CDM cosmology using residuals. Our analysis indicates that the preference for $\xi<0$ is primarily driven by the high multipoles data. For instance, in both ACT and ACT+Planck, this improvement is due to the lower amplitude of the ACT acoustic modes at high $\ell$, which represents a genuine effect rather than the fake one discussed in Ref.~\cite{DiValentino:2020leo} for Planck data alone. Future accurate measurements of the CMB damping tail can shed much light not only on IDE cosmologies but also in extended $\Lambda$CDM scenarios (such as those with extra relativistic degrees of freedom) making therefore stronger the case for these future CMB probes, as the CMB stage IV mission~\cite{CMB-S4:2022ght}.

\acknowledgments
\noindent  
CvdB is supported (in part) by the Lancaster–Manchester–Sheffield Consortium for Fundamental Physics under STFC grant: ST/T001038/1. EDV is supported by a Royal Society Dorothy Hodgkin Research Fellowship. RCN thanks the CNPq for partial financial support under the project No. 304306/2022-3.
This article is based upon work from COST Action CA21136 Addressing observational tensions in cosmology with systematics and fundamental physics (CosmoVerse) supported by COST (European Cooperation in Science and Technology). We acknowledge IT Services at The University of Sheffield for the provision of services for High Performance Computing. 
This work has been partially supported by the MCIN/AEI/10.13039/501100011033 of Spain under grant PID2020-113644GB-I00, by the Generalitat Valenciana of Spain under the grant PROMETEO/2019/083 and by the European Union’s Framework Programme for Research and Innovation Horizon 2020 (2014–2020) under grant H2020-MSCA-ITN-2019/860881-HIDDeN.

%\begin{thebibliography}
\bibliographystyle{unsrt}
\bibliography{biblio}
%\end{thebibliography}

% Please avoid comments such as "For a review'', "For some examples",
% "and references therein" or move them in the text. In general,
% please leave only references in the bibliography and move all
% accessory text in footnotes.

% Also, please have only one work for each \bibitem.

\end{document}